\documentclass[aps,showpacs,pra,superscriptaddress,]{revtex4}
\usepackage{amsmath}
\usepackage{amsfonts}
\usepackage{amssymb}
\usepackage{bm}
\usepackage{graphicx}

\begin{document}

\title{Generation of solitons and periodic wave trains in birefringent
optical fibers }
\author{ Houria Triki}
\affiliation{Radiation Physics Laboratory, Department of Physics, Faculty of Sciences,
Badji Mokhtar University, P. O. Box 12, 23000 Annaba, Algeria}
\author{Vladimir I. Kruglov}
\affiliation{Centre for Engineering Quantum Systems, School of Mathematics and Physics,
The University of Queensland, Brisbane, Queensland 4072, Australia}

\begin{abstract}
We investigate the existence and propagation properties of all possible
types of envelope soliton pulses in a birefringent optical fiber wherein the
light propagation is governed by two coupled nonlinear Schr\"{o}dinger
equations with coherent and incoherent nonlinear couplings. Especially, we
study the existence of optical solitons under the influence of
group-velocity dispersion and third-order nonlinearity, which have physical
relevance in the context of elliptical core optical fiber. The results show
that the waveguiding medium supports the existence of a wide variety of
propagating envelope solitons, including dipole-bright, bright-dipole,
bright-dark, dark-bright, W-shaped-dipole and dipole-W-shaped pulses which
exhibit different characteristics. Interestingly, we find that the obtained
soliton pairs are allowable in both the normal and anomalous dispersion
regimes. A wide class of exact analytic periodic (elliptic) wave solutions are identified. This illustrates the potentially rich set of localized
pulses and nonlinear periodic waves in birefringent optical fiber media.
\end{abstract}

\maketitle
\affiliation{}

\section{Introduction}

Propagation dynamics of envelope solitons in birefringent optical fibers has
been the subject of intensive experimental and theoretical studies in recent
years \cite{R1,R2}. This because of their important potential applications
for the design of new kinds of all-optical switches and logic gates \cite%
{Islam,Wai}. One should note here that no mode is single in realistic
nonlinear fibers due to the presence of birefringence phenomenon \cite%
{Menyuk} and single mode fibers are normally bimodal \cite{Triki}.
Interestingly, the study of electromagnetic field dynamics in birefringent
optical media showed a multitude of new behaviors not demonstrated in the
scalar case, for example, by exploring the concept of a vector soliton. The
principal physical processes that are responsible for the soliton formation
in such birefringent fibers are the group-velocity dispersion, self-phase
modulation and cross-phase modulation. In this situation, the main nonlinear
evolution equations governing the light propagation in such transmission
systems are the coupled nonlinear Schr\"{o}dinger (NLS) equations which
constitute the well known Manakov system \cite{M1,M2}. Such integrable
system represents a natural and important extension of the scalar cubic NLS
equation, which describe the interaction between distinct field components.
It is worthy to note that the relevance of coupled Manakov-type equations
for describing vector phenomena is not only restricted to optics \cite{R3},
but also to Bose-Einstein condensates \cite{R4}, hydrodynamics \cite{R5},
and plasma physics \cite{R6}. Importantly, Manakov vector solitons in
nonlinear systems can take different shapes and forms, including
bright-dark, bright-bright, and dark-dark solitons \cite{R7,R8,R9}. It is of
interest to note that Manakov solitons have been demonstrated experimentally
in real physical media and natural environments such as photorefractive
crystals \cite{A1,A2}, optical fiber \cite{A3}, quadratic media \cite{A4},
semiconductor waveguides \cite{A5}, and Bose-Einstein condensates \cite{A6}.

In reality, the simultaneous propagation of multiple light pulses or beams
in nonlinear systems is modeled by coupled NLS-like equations which are
non-integrable in general \cite{Kanna}. These multicomponent NLS equations
can be divided into two categories, namely coherently and incoherently
coupled NLS equations \cite{R3}. This coupled form of NLS equations plays a
crucial role in optics due to their important applications in multichannel
bit parallel-wavelength and wavelength-division multiplexing optical fiber
networks \cite{R10,R11,R12}.

Presently, considerable interest is still paid to search for new kinds of
solitons/solitary waves, uncover the optical transmission and stability
properties of them, and understand the mechanism of them. This is not
surprising because these localized structures have a broad range of
practical applications encompassing engineering and nonlinear science \cite%
{R3,Akhmediev}. We should note here that solutions in an exact analytical
form are highly desired when one compare experimental findings with theory.
Besides, closed form solutions when they exist can enable one to determine
some relevant physical quantities analytically and find results of numerical
simulations. In this paper, we show that a\ birefringent optical fiber
system can support a different variety of interesting soliton pulses, which
have different characteristics (e.g., inverse temporal width, amplitudes,
and wave numbers) but propagate with identical group velocities.
Importantly, the derived exact soliton solutions contain free parameters,
thus implying that one can finds families of propagating localized waves in
the material.

The paper is organized as follows. In Sec. II, we present the coupled NLS
system with coherent and incoherent nonlinear couplings describing the
evolution of two field components in a birefringent optical fiber and
give a detailed derivation of its exact analytical solitons solutions. In
paraticular, classes of soliton pair solutions in the form of dipole-bright,
bright-dipole, bright-dark, dark-bright, W-shaped-dipole, and
dipole-W-shaped solitons solutions are obtained here. The characteristics of
these soliton pulses are also discussed. Various periodic wave solutions for
this model, which are expressed in terms of the Jacobian elliptic functions
are derived in Sec. III. Finally, our conclusions are given in Sec. IV.

\section{Coupled NLS system and its exact soliton solutions}

\subsection{Theoretical model}

We consider here the coupled-mode theory for directional couplers. The NLS
equations simplify considerably for a symmetric coupler with two identical
cores. In this case the coupled-mode equations for symmetric couplers become 
\cite{R11}: 
\begin{equation}
i\frac{\partial \psi _{1}}{\partial z}-\frac{\beta _{2}}{2}\frac{\partial
^{2}\psi _{1}}{\partial \tau ^{2}}+\gamma (\left\vert \psi _{1}\right\vert
^{2}+\sigma \left\vert \psi _{2}\right\vert ^{2})\psi _{1}+\alpha \psi
_{2}=0,  \label{1}
\end{equation}%
\begin{equation}
i\frac{\partial \psi _{2}}{\partial z}-\frac{\beta _{2}}{2}\frac{\partial
^{2}\psi _{2}}{\partial \tau ^{2}}+\gamma (\left\vert \psi _{2}\right\vert
^{2}+\sigma \left\vert \psi _{1}\right\vert ^{2})\psi _{2}+\alpha \psi
_{1}=0,  \label{2}
\end{equation}%
where $\psi _{1}$ and $\psi _{2}$ represent the slowly varying envelopes of
the electromagnetic waves, $z$ and $\tau $ are the propagation distance and
reduced time $\tau =t-\beta _{1}z$. Here $\beta _{1}$, $\beta _{2}$, $\gamma 
$, $\sigma $ and $\alpha $ are dispersions, self- and cross-phase
modulations, and coupling coefficient respectively. We introduce two new
variables as 
\begin{equation}
U=\frac{1}{\sqrt{2}}(\psi _{1}+\psi _{2}),\quad V=\frac{1}{\sqrt{2}}(\psi
_{1}-\psi _{2}),  \label{3}
\end{equation}%
which associated with the even and odd supermodes. In terms of new complex
amplitudes $U$ and $V,$ Eqs. (\ref{1}) and (\ref{2}) have the following
form, 
\begin{equation}
i\frac{\partial U}{\partial z}-\frac{\beta _{2}}{2}\frac{\partial ^{2}U}{%
\partial \tau ^{2}}+(\gamma _{1}\left\vert U\right\vert ^{2}+\gamma
\left\vert V\right\vert ^{2})U+\gamma _{2}V^{2}U^{\ast }+\alpha U=0,
\label{4}
\end{equation}%
\begin{equation}
i\frac{\partial V}{\partial z}-\frac{\beta _{2}}{2}\frac{\partial ^{2}V}{%
\partial \tau ^{2}}+(\gamma _{1}\left\vert V\right\vert ^{2}+\gamma
\left\vert U\right\vert ^{2})V+\gamma _{2}U^{2}V^{\ast }-\alpha V=0,
\label{5}
\end{equation}%
where $\gamma _{1}=\gamma (1+\sigma )/2$ and $\gamma _{2}=\gamma (1-\sigma
)/2$. In the what follows, we are interested by waveform solutions of Eqs. (\ref{4}) and (\ref{5}) which exhibit different localized profile of
amplitudes. The existence of such nonlinear waves is of practical relevance
as they describe the most general forms of traveling waves with different
properties. The complex envelope traveling-waves $U(z,\tau )$ and $V(z,\tau) $ can be written \cite{Kruglov1,Kruglov2} as
\begin{equation}
U(z,\tau )=u(x)\exp [i(\kappa z-\delta \tau +\theta )],  \label{6}
\end{equation}%
\begin{equation}
V(z,\tau )=v(x)\exp [i(\kappa z-\delta \tau +\theta )],  \label{7}
\end{equation}%
where $u(x)$ and $v(x)$ are real amplitude functions depending on the
traveling coordinate $x=\tau -qz$, and $q=1/\mathrm{v}$ is the inverse
velocity of the waves. Also, $\kappa $ and $\delta $ are the respective real
parameters describing the wave number and frequency shift, while $\theta $
represent the phase of pulses at $z=0$.

On substituting the waveform solutions (\ref{6}) and (\ref{7}) into Eqs. (\ref{4}) and (\ref{5}) and separating the real and imaginary parts, one
obtains the two ordinary differential equations: 
\begin{equation}
\frac{\beta _{2}}{2}\frac{d^{2}u}{dx^{2}}-\gamma _{1}u^{3}-(\gamma +\gamma
_{2})v^{2}u+\left( \kappa -\alpha -\frac{1}{2}\beta _{2}\delta ^{2}\right)
u=0,  \label{8}
\end{equation}%
\begin{equation}
\frac{\beta _{2}}{2}\frac{d^{2}v}{dx^{2}}-\gamma _{1}v^{3}-(\gamma +\gamma
_{2})u^{2}v+\left( \kappa +\alpha -\frac{1}{2}\beta _{2}\delta ^{2}\right)
v=0.  \label{9}
\end{equation}%
We have also found the relation $q=\beta _{2}\delta $ showing that the
inverse velocity $q$ is controlled by $\delta $. One can express Eqs. (\ref{8}) and (\ref{9}) in the form, 
\begin{equation}
\frac{d^{2}u}{dx^{2}}+au^{3}+bv^{2}u+cu=0,  \label{10}
\end{equation}%
\begin{equation}
\frac{d^{2}v}{dx^{2}}+av^{3}+bu^{2}v+dv=0,  \label{11}
\end{equation}%
where the parameters $a$, $b$, $c$ and $d$ are given by%
\begin{equation}
a=-\frac{\gamma (1+\sigma )}{\beta _{2}},\quad b=\frac{\gamma (\sigma -3)}{%
\beta _{2}},\quad c=\frac{2(\kappa -\alpha )}{\beta _{2}}-\delta ^{2},\quad
d=\frac{2(\kappa +\alpha )}{\beta _{2}}-\delta ^{2}.  \label{12}
\end{equation}

In what follows, we consider the case when $\sigma $ $=1$, which corresponds
to a strong birefringent effect of the fiber. This leads to $a=b$ in Eq. (%
\ref{12}) and consequently the amplitude Eqs. (\ref{10}) and (\ref{11})
become as
\begin{equation}
\frac{d^{2}u}{dx^{2}}+au^{3}+av^{2}u+cu=0,  \label{13}
\end{equation}%
\begin{equation}
\frac{d^{2}v}{dx^{2}}+av^{3}+au^{2}v+dv=0,  \label{14}
\end{equation}

\noindent where%
\begin{equation}
a=-\frac{2\gamma }{\beta _{2}},\quad c=\frac{2(\kappa -\alpha )}{\beta _{2}}%
-\delta ^{2},\quad d=\frac{2(\kappa +\alpha )}{\beta _{2}}-\delta ^{2}.
\label{15}
\end{equation}

\subsection{Envelope solitons}

Our purpose now is to analyze the existence of exact analytical soliton
solutions of the coupled NLS equations (\ref{1}) and (\ref{2}) based on
solving the amplitudes Eqs. (\ref{13}) and (\ref{14}). In what follows,
various kinds of localized wave packets with interesting properties are
presented, which may have potential applications in birefringent fiber
systems.

\subsubsection{Dipole-bright solitons}

The first type of localized solutions we obtained here for the amplitude
equations (\ref{13}) and (\ref{14}) consists of dipole and bright solitons
which take the form,%
\begin{equation}
u\left( x\right) =\lambda \,\mathrm{sech}[w(x-\eta )]\mathrm{tanh}\left[
w(x-\eta )\right] ,  \label{16}
\end{equation}%
\begin{equation}
v\left( x\right) =\rho \,\mathrm{sech}^{2}[w(x-\eta )],  \label{17}
\end{equation}%
where%
\begin{equation}
w^{2}=-c,\quad \lambda ^{2}=-\frac{6c}{a},\quad \rho^{2} =\lambda^{2} ,\quad
d=4c.  \label{18}
\end{equation}

\noindent and $\eta $ being the position of the nonlinear wavepackets at $z=0 $.

Then, using Eqs. (\ref{15}) and (\ref{18}), we get following relations for
the inverse temporal width $w$, amplitudes $\lambda $ and wave number $\kappa $: 
\begin{equation}
w=\sqrt{-\frac{4\alpha }{3\beta _{2}}},\quad \lambda =\pm \sqrt{\frac{%
4\alpha }{\gamma }},\quad \kappa =\frac{5\alpha }{3}+\frac{\beta _{2}\delta
^{2}}{2},  \label{19}
\end{equation}%
where $\delta$ is free parameter. We have for positive and negative value of
parameter $\lambda $ two different solution for the amplitude $\rho $ as 
\begin{equation}
\rho =\sqrt{\frac{4\alpha }{\gamma }},\quad \rho =-\sqrt{\frac{4\alpha }{%
\gamma }},  \label{20}
\end{equation}%
because of relation $\rho ^{2}=\lambda ^{2}$. Substitution of these results
into Eqs. (\ref{6}) and (\ref{7}), one gets a class of exact soliton
solutions for the coupled NLS system (\ref{4}) and (\ref{5}) as
\begin{equation}
U(z,\tau )=\lambda \,\mathrm{sech}[w(\tau -qz-\eta )]\mathrm{tanh}\left[
w(\tau -qz-\eta )\right] \exp [i(\kappa z-\delta \tau +\theta )],  \label{21}
\end{equation}

\begin{equation}
V(z,\tau )=\rho \,\mathrm{sech}^{2}[w(\tau -qz-\eta )]\exp [i(\kappa
z-\delta \tau +\theta )],  \label{22}
\end{equation}%
which exist for waveguide parameters obeying the physical conditions $\alpha
\beta _{2}<0$ and $\alpha \gamma >0$. The wave number $\kappa $ in soliton
solutions (\ref{21}) and (\ref{22}) depends on a free parameter $\delta $.

\begin{figure}[h]
\includegraphics[width=1\textwidth]{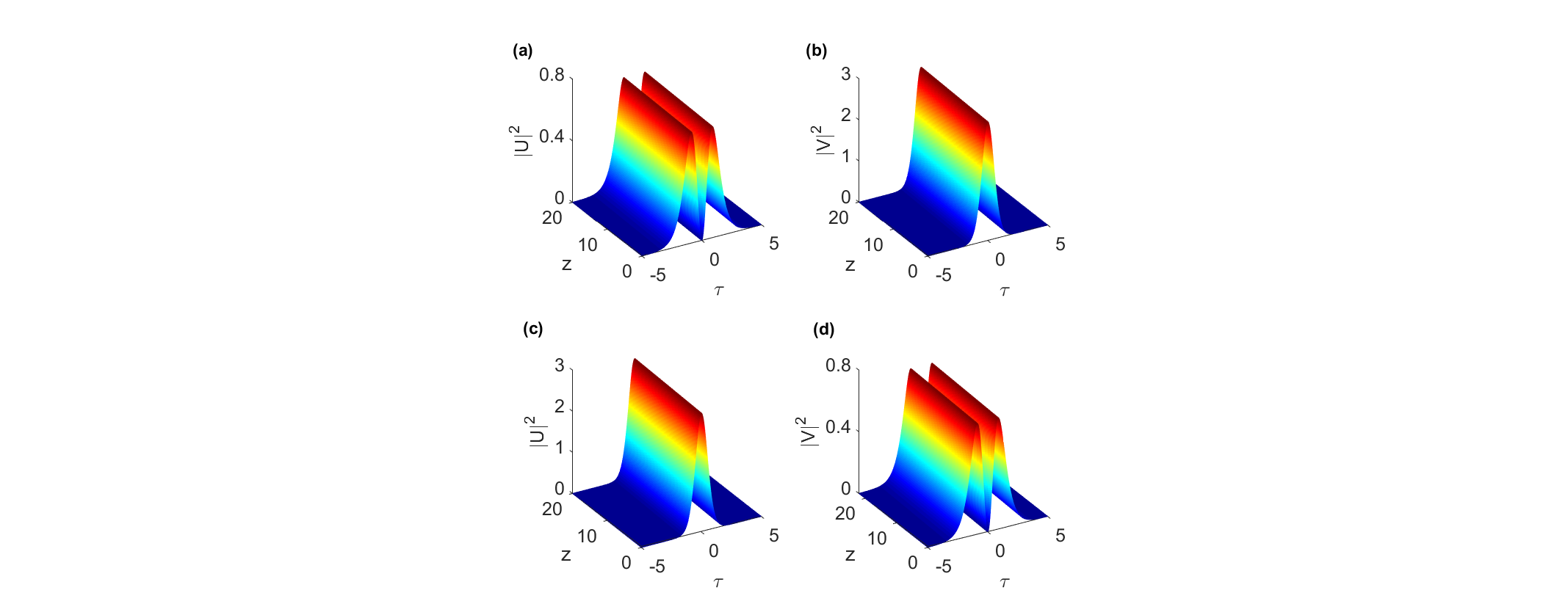}
\caption{Intensity profiles of (a) dipole soliton (\protect\ref{21}) (b)
bright soliton (\protect\ref{22}) (c) bright soliton (\protect\ref{28}) and
(d) dipole soliton (\protect\ref{29}) for the values mentioned in the text.}
\label{FIG.1.}
\end{figure}

An example of the soliton solutions (\ref{21}) and (\ref{22}) is presented
in Figs. 1(a) and (b), using the parameter values $\beta _{2}=-1,\alpha
=0.75,$ $\gamma =1,$ $\delta =-0.01,$ and $\eta =0$. It is clear from 
figures that these localized solutions describe dipole-bright waveforms in
the birefringent optical fiber. As seen from the expressions (\ref{19}) and (%
\ref{20}), the widths and amplitudes of solitons depend on the fiber
parameters. This allow us to obtain the desired soliton power and width
based on an appropriate choice of the fiber characteristic parameters.

\subsubsection{Bright-dipole solitons}

We have found the closed form solutions for the amplitude equations (\ref{13}%
) and (\ref{14})\ in the form of bright-dipole solitons as 
\begin{equation}
u\left( x\right) =R\,\mathrm{sech}^{2}[\mu (x-\eta )],  \label{23}
\end{equation}%
\begin{equation}
v\left( x\right) =P\,\mathrm{sech}[\mu (x-\eta )]\mathrm{tanh}\left[ \mu
(x-\eta )\right] ,  \label{24}
\end{equation}%
where%
\begin{equation}
\mu ^{2}=-d,\quad P^{2}=-\frac{6d}{a},\quad R^{2}=P^{2},\quad d=%
\frac{c}{4}.  \label{25}
\end{equation}

\noindent Equations (\ref{15}) and (\ref{25}) yield the soliton parameters:%
\begin{equation}
\mu =\sqrt{\frac{4\alpha }{3\beta _{2}}},\quad P=\pm \sqrt{-\frac{4\alpha }{%
\gamma }},\quad \kappa =-\frac{5\alpha }{3}+\frac{\beta _{2}\delta ^{2}}{2},
\label{26}
\end{equation}%
where $\delta $ is free parameter. We have for positive and negative value
of parameter $P$ two different solution for the amplitude $R$ as 
\begin{equation}
R=\sqrt{-\frac{4\alpha }{\gamma }},\quad R=-\sqrt{-\frac{4\alpha }{\gamma }},
\label{27}
\end{equation}%
because of relation $R^{2}=P^{2}$. \noindent If we insert these results into
Eqs. (\ref{6}) and (\ref{7}), one obtains exact analytical soliton solutions
for the coupled NLS equations (\ref{4}) and (\ref{5}) as
\begin{equation}
U(z,\tau )=R\,\mathrm{sech}^{2}[\mu (\tau -qz-\eta )]\exp [i(\kappa z-\delta
\tau +\theta )],  \label{28}
\end{equation}

\begin{equation}
V(z,\tau )=P\,\mathrm{sech}[\mu (\tau -qz-\eta )]\mathrm{tanh}\left[ \mu
(\tau -qz-\eta )\right] \exp [i(\kappa z-\delta \tau +\theta )].  \label{29}
\end{equation}

We observe that the existence conditions of bright-dipole are $\alpha \beta
_{2}>0$ and $\alpha \gamma <0,$ which differ from those for dipole-bright
solitons. On comparing Eqs. (\ref{19}) and (\ref{20}) to (\ref{26}) and (\ref%
{27}), we find that the inverse temporal width, amplitudes and wavenumber of
bright-dipole solitons are different from those of dipole-bright solitons.

The intensity profiles of $\mathrm{sech}^{2}$-type bright and dipole
solitons (\ref{28}) and (\ref{29}) are presented in Figs. 1(c) and (d),
respectively. Here we have taken the parameters values $\beta _{2}=-1,\alpha
=-0.75,$ $\gamma =1,$ $\delta =0.01,$ and $\eta =0$. These solitons have
amplitudes and width which are completely determined by the fiber parameters
while their wave number is related to the free parameter $\delta .$

It is interesting to note that dipole soliton pulses have been obtained in
single mode fibers with non-Kerr nonlinearity, for which the formation of
this kind of pulse shape requires the compensation of both second- and
third-order dispersion effects \cite{Porsezian}. For the above dipole pulses
no compensation of dispersion is needed for their generation, which may 
be advantageous in various practical applications. We also note that
controllable dipole-mode solitons for generated four-wave mixing have been
experimentally demonstrated in an optically induced atomic system \cite{YZhang}. The occurrence of dipole and W-shaped soliton pulses in the
birefringent fiber make the current optical system ideal to study the
formations of multi-humped solitons and their nonlinear dynamics.

\subsubsection{\noindent Bright-dark solitons}

We find that Eqs. (\ref{13}) and (\ref{14}) admit exact analytical soliton
solutions of the form,%
\begin{equation}
u\left( x\right) =S\,\mathrm{sech}\left[ \sigma (x-\eta )\right] ,
\label{30}
\end{equation}%
\begin{equation}
v\left( x\right) =Q\,\mathrm{tanh}\left[ \sigma (x-\eta )\right] ,
\label{31}
\end{equation}%
where%
\begin{equation}
\sigma ^{2}=d-c,\quad S^{2}=\frac{d-2c}{a},\quad Q^{2}=-\frac{d}{a}.
\label{32}
\end{equation}
As follows from Eqs. (\ref{15}) and (\ref{32}), the soliton parameters $\sigma $ and $S$ are 
\begin{equation}
\sigma =\sqrt{\frac{4\alpha }{\beta _{2}}},  \label{33}
\end{equation}%
\begin{equation}
S=\pm \sqrt{\frac{2\kappa -\beta _{2}\delta ^{2}-6\alpha }{2\gamma }}.
\label{34}
\end{equation}%
We have for positive and negative value of parameter $S$ two different
solution for the amplitude $Q$ as 
\begin{equation}
Q=\sqrt{\frac{2\kappa -\beta _{2}\delta ^{2}+2\alpha }{2\gamma }},\quad Q=-%
\sqrt{\frac{2\kappa -\beta _{2}\delta ^{2}+2\alpha }{2\gamma }}.  \label{35}
\end{equation}%
Substitution of these waveform solutions into Eqs. (\ref{6}) and (\ref{7})
yield the following exact soliton solutions for the coupled NLS system (\ref%
{4}) and (\ref{5}):%
\begin{equation}
U(z,\tau )=S\,\mathrm{sech}\left[ \sigma (\tau -qz-\eta )\right] \exp
[i(\kappa z-\delta \tau +\theta )],  \label{36}
\end{equation}%
\begin{equation}
V(z,\tau )=Q\,\mathrm{tanh}\left[ \sigma (\tau -qz-\eta )\right] \exp
[i(\kappa z-\delta \tau +\theta )].  \label{37}
\end{equation}

\begin{figure}[h]
\includegraphics[width=1\textwidth]{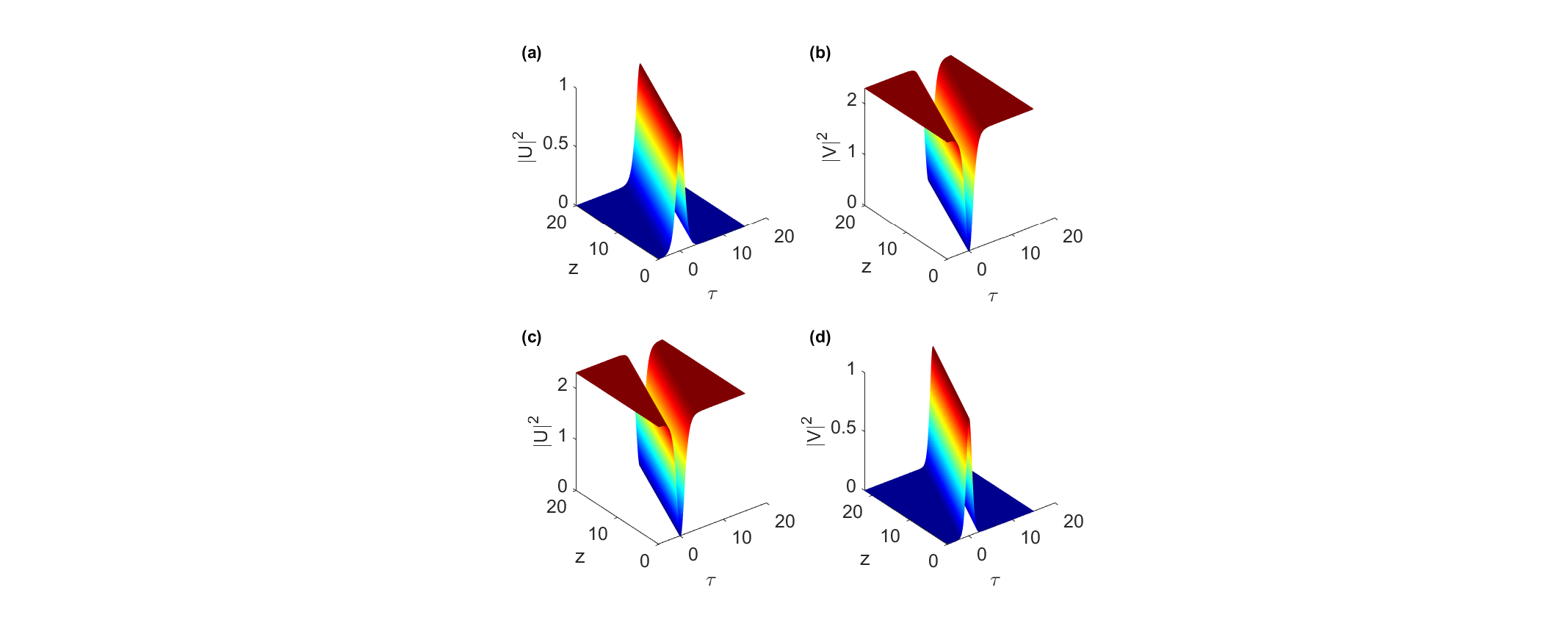}
\caption{Intensity profiles of (a) bright soliton (\protect\ref{36}) (b)
dark soliton (\protect\ref{37}) (c) dark soliton (\protect\ref{44}) and (d)
bright soliton (\protect\ref{45}) for the values mentioned in the text.}
\label{FIG.2.}
\end{figure}

Note that these soliton solutions are dependent on two free parameters $\kappa $ and $\delta $ and exist for $\alpha \beta _{2}>0$, $\gamma (2\kappa
-\beta _{2}\delta ^{2}-6\alpha )>0$ and $\gamma (2\kappa -\beta _{2}\delta^{2}+2\alpha )>0$. The intensity profiles of these bright- and dark-type
solitons are depicted in Figs. 2(a) and (b), respectively, for $\beta_{2}=1, \alpha =\frac{1}{6},$ $\gamma =0.5,$ $\delta =0.5,$ $\kappa =1,125$
and $\eta =0$. These solitons have different amplitudes and consequently different powers.

\subsubsection{Dark-bright solitons}

We obtained a class of exact analytical soliton solutions for Eqs. (\ref{13}) and (\ref{14}) as
\begin{equation}
u\left( x\right) =f\,\mathrm{tanh}\left[ r(x-\eta )\right] ,  \label{38}
\end{equation}%
\begin{equation}
v\left( x\right) =g\,\mathrm{sech}\left[ r(x-\eta )\right] ,  \label{39}
\end{equation}%
where%
\begin{equation}
r^{2}=c-d\quad f^{2}=-\frac{c}{a},\quad g^{2}=\frac{c-2d}{a}.  \label{40}
\end{equation}

\noindent Hence, using Eqs. (\ref{15}) and (\ref{40}), we obtain the soliton
parameters $r$ and $f$ as 
\begin{equation}
r=\sqrt{-\frac{4\alpha }{\beta _{2}}},  \label{41}
\end{equation}%
\begin{equation}
f=\pm \sqrt{\frac{2\kappa -\beta _{2}\delta ^{2}-2\alpha }{2\gamma }}.
\label{42}
\end{equation}%
We have for positive and negative value of parameter $f$ two different
solution for the amplitude $g$ as 
\begin{equation}
g=\sqrt{\frac{2\kappa -\beta _{2}\delta ^{2}+6\alpha }{2\gamma }},\quad g=-%
\sqrt{\frac{2\kappa -\beta _{2}\delta ^{2}+6\alpha }{2\gamma }}.  \label{43}
\end{equation}%
Using these expressions, we can write the exact dark-bright soliton
solutions of the coupled NLS equations (\ref{4}) and (\ref{5}) as
\begin{equation}
U(z,\tau )=f\,\mathrm{tanh}\left[ r(\tau -qz-\eta )\right] \exp [i(\kappa
z-\delta \tau +\theta )],  \label{44}
\end{equation}%
\begin{equation}
V(z,\tau )=g\,\mathrm{sech}\left[ r(\tau -qz-\eta )\right] \exp [i(\kappa
z-\delta \tau +\theta )].  \label{45}
\end{equation}%
Note that these soliton solutions are dependent on two free parameters $\kappa $ and $\delta $, and they do exist provided that $\alpha \beta _{2}<0,$
$\gamma (2\kappa -\beta _{2}\delta ^{2}-2\alpha )>0$ and $\gamma (2\kappa
-\beta _{2}\delta ^{2}+6\alpha )>0$. It is clear from Eqs. (\ref{41}), (\ref%
{42}) and (\ref{43}) that the inverse temporal width and amplitudes of
dark-bright solitons are different from those obtained for bright-dark
solitons in Eqs. (\ref{33}), (\ref{34}) and (\ref{35}). In Figs. 2(c) and
(d), we have presented the intensity of these dark and bright solitons for
the values $\beta _{2}=1,\alpha =-\frac{1}{6},$ $\gamma =0.5,$ $\delta =0.5,$
$\kappa =1,125$, and $\eta =0$. The above results show that the amplitude
and width of both bright-dark and dark-bright solitons are completely
determined by two free parameters, the values of wave number $\kappa $ and the frequency shift parameter $\delta $. This means that families of this kind of localized waves can exist in the birefringent optical medium.

\subsubsection{W-shaped-dipole solitons}

An interesting class of exact analytical soliton solutions can be obtained
for Eqs. (\ref{13}) and (\ref{14}) as 
\begin{equation}
u(x)=A-B\,\mathrm{sech}^{2}[\epsilon (x-\eta )],  \label{46}
\end{equation}%
\begin{equation}
v(x)=F\,\mathrm{sech}[\epsilon (x-\eta )]\mathrm{tanh}\left[ \epsilon
(x-\eta )\right] .  \label{47}
\end{equation}%
The parameters in these solutions are given by equations: $A^{2}=-c/a$, $%
B^{2}=-9c/4a$, $F^{2}=B^{2}$, $\epsilon ^{2}=c/8$ and $d=7c/8$. Thus, we
have the following equations for the amplitudes as 
\begin{equation}
A=\pm \sqrt{-\frac{c}{a}},\quad B=\pm \frac{3}{2}\sqrt{-\frac{c}{a}},\quad
F=\pm \frac{3}{2}\sqrt{-\frac{c}{a}}.  \label{48}
\end{equation}%
We note that the signs for amplitudes in (\ref{48}) are independent. Hence,
one can choose eight different sequences of signs in (\ref{48}). This means
we have eight different solutions given by Eqs. (\ref{46}) and (\ref{47}).
We also have the following two equations as 
\begin{equation}
\epsilon =\frac{1}{2}\sqrt{\frac{c}{2}},\quad d=\frac{7}{8}c,  \label{49}
\end{equation}%
where without loss of generality we choose the positive sign for parameter $%
\epsilon $. Using Eqs. (\ref{15}), (\ref{48}) and (\ref{49}), we get the
following soliton parameters: 
\begin{eqnarray}
&&\left. A=\pm 4\sqrt{-\frac{\alpha }{\gamma }},\quad B=\pm 6\sqrt{-\frac{%
\alpha }{\gamma }},\quad F=\pm 6\sqrt{-\frac{\alpha }{\gamma }},\right.
\label{50} \\
&&\left. \epsilon =2\sqrt{-\frac{\alpha }{\beta _{2}}}\right. ,  \label{51}
\\
&&\left. \kappa =-15\alpha +\frac{1}{2}\beta _{2}\delta ^{2}.\right.
\label{52}
\end{eqnarray}

\begin{figure}[h]
\includegraphics[width=1\textwidth]{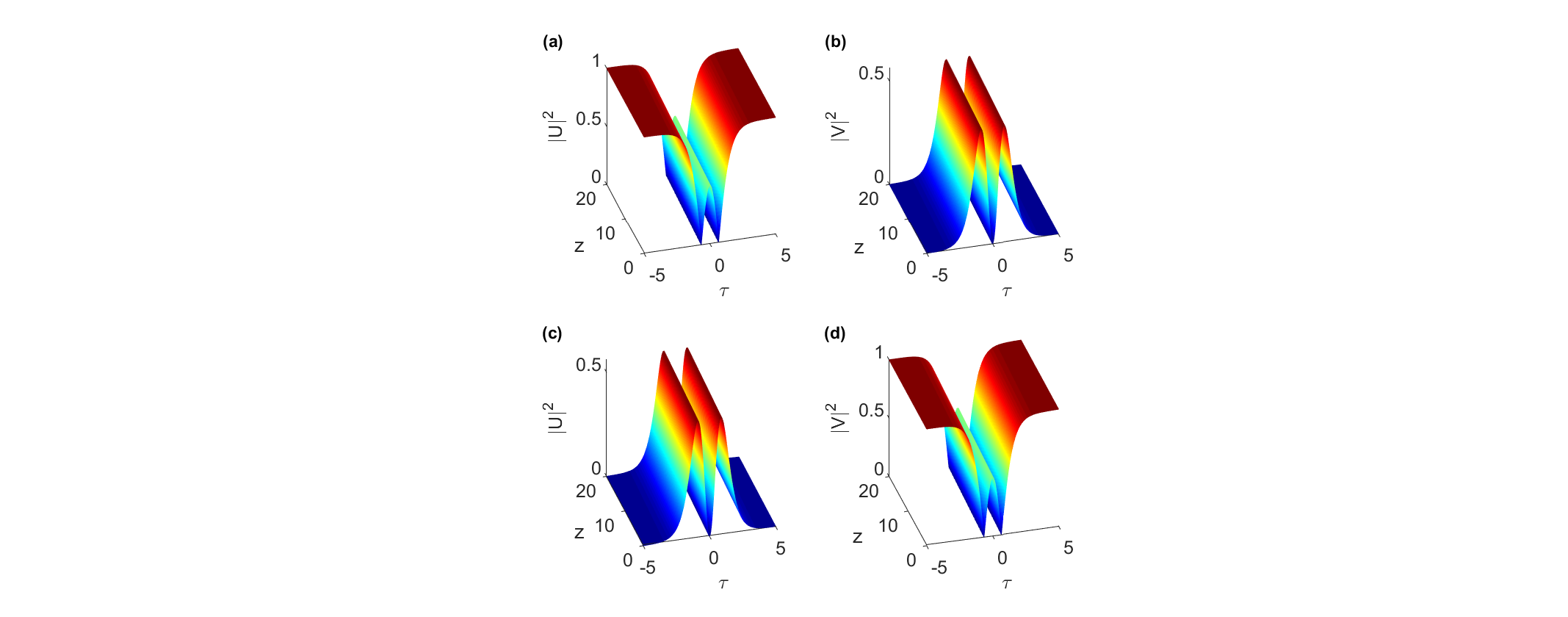}
\caption{Intensity profiles of (a) W-shaped soliton (\protect\ref{53}) (b)
dipole soliton (\protect\ref{54}) (c) dipole soliton (\protect\ref{62}) and
(d) W-shaped soliton (\protect\ref{63}) for the values mentioned in the
text. }
\label{FIG.3.}
\end{figure}

Note that relation for the wave number $\kappa $ in (\ref{52}) with an
arbitrary parameter $\delta $ follows from equation $d=7c/8$. We can choose
eight different sequences of signs in (\ref{50}) which yields eight
different solutions given by Eqs. (\ref{46}) and (\ref{47}). The
substitution of these results into Eqs. (\ref{6}) and (\ref{7}) yields eight
soliton pulse solutions for the coupled NLS equations (\ref{4}) and (\ref{5}%
) as 
\begin{equation}
U(z,\tau )=\left\{ A-B\mathrm{sech}^{2}[\epsilon (\tau -qz-\eta )]\right\}
\exp [i(\kappa z-\delta \tau +\theta )],  \label{53}
\end{equation}%
\begin{equation}
V(z,\tau )=F\,\mathrm{sech}[\epsilon (\tau -qz-\eta )]\mathrm{tanh}\left[
\epsilon (\tau -qz-\eta )\right] \exp [i(\kappa z-\delta \tau +\theta )],
\label{54}
\end{equation}
which exist provided that $\alpha \gamma <0$ and $\alpha \beta _{2}<0$.

The intensity profile for these solutions are shown in Figs. 3(a) and (b)\ when $\beta _{2}=0.5,\alpha =-0.125,$ $\gamma =2,$ $\delta =0.02,$ and $\eta =0$.
These figures demonstrate that the pulse envelopes $U$ and $V$ represent
W-shaped and dipole solitons respectively. The fact that these soliton
structures involve more than one intensity peak (multihump) is very
intriguing, as they are very rarely obtainable in the coupled NLS equations
framework.

\subsubsection{Dipole-W-shaped solitons}

We obtained another class of exact analytical soliton solutions for Eqs. (\ref{13}) and (\ref{14}) as 
\begin{equation}
u\left( x\right) =\Lambda \,\mathrm{sech}[\nu (x-\eta )]\mathrm{tanh}\left[
\nu (x-\eta )\right] ,  \label{55}
\end{equation}%
\begin{equation}
v\left( x\right) =D-\Gamma \,\mathrm{sech}^{2}[\nu (x-\eta )].  \label{56}
\end{equation}%
The parameters $D$, $\Gamma $ and $\Lambda $ are 
\begin{equation}
D=\pm \sqrt{-\frac{d}{a}},\quad \Gamma =\pm \frac{3}{2}\sqrt{-\frac{d}{a}}%
,\quad \Lambda \,=\pm \frac{3}{2}\sqrt{-\frac{d}{a}}.  \label{57}
\end{equation}%
We note that the signs for amplitudes in (\ref{57}) are independent. Hence,
one can choose eight different sequences of signs in (\ref{57}). This means
we have eight different solutions given by Eqs. (\ref{55}) and (\ref{56}).
We also have the following two equations as 
\begin{equation}
\nu =\frac{1}{2}\sqrt{\frac{d}{2}},\quad c=\frac{7}{8}d.  \label{58}
\end{equation}%
Using Eqs. (\ref{15}), (\ref{57}) and (\ref{58}) we get the soliton
parameters: 
\begin{eqnarray}
&&\left. D=\pm 4\sqrt{\frac{\alpha }{\gamma }},\quad \Gamma =\pm 6\sqrt{%
\frac{\alpha }{\gamma }},\quad \Lambda =\pm 6\sqrt{\frac{\alpha }{\gamma }}%
,\right.  \label{59} \\
&&\left. \nu =2\sqrt{\frac{\alpha }{\beta _{2}}}\right. ,  \label{60} \\
&&\left. \kappa =15\alpha +\frac{1}{2}\beta _{2}\delta ^{2}.\right.
\label{61}
\end{eqnarray}%
Note that relation for the wave number $\kappa $ in (\ref{61}) with an
arbitrary parameter $\delta $ follows from equation $c=7d/8$. We can choose
eight different sequences of signs in (\ref{59}) which yields eight
different solutions given by Eqs. (\ref{55}) and (\ref{56}). The substitution of these results into Eqs. (\ref{6}) and (\ref{7}) yields eight
soliton pulse solutions for the coupled NLS equations (\ref{4}) and (\ref{5}) as 
\begin{equation}
U(z,\tau )=\Lambda \,\mathrm{sech}[\nu (\tau -qz-\eta )]\mathrm{tanh}\left[
\nu (\tau -qz-\eta )\right] \exp [i(\kappa z-\delta \tau +\theta )],
\label{62}
\end{equation}%
\begin{equation}
V(z,\tau )=\left\{ D-\Gamma \,\mathrm{sech}^{2}[\nu (\tau -qz-\eta
)]\right\} \exp [i(\kappa z-\delta \tau +\theta )],  \label{63}
\end{equation}

\noindent which exist provided that $\alpha \gamma >0$ and $\alpha \beta
_{2}>0$. Figures 3(c) and (d) depict the intensity profiles of such dipole
and W-shaped solitons for the values $\beta _{2}=0.5,\alpha =0.125,$ $\gamma
=2,$ $\delta =0.02,$ and $\eta =0$.

Before ending this section, we give a useful remark on the influence of the
group-velocity dispersion on soliton properties. We find from the above
results that the existence of all the soliton types is related to the sign
of the product $\left( \alpha \beta _{2}\right) $. Hence, depending on the
choice of the value of coupling coefficient $\alpha $, these soliton pairs
are obtainable in both the normal ($\beta _{2}>0$) and anomalous ($\beta_{2}<0$) dispersion regimes. Now looking closely to the amplitudes and
widths of bright-dipole, dipole-bright, dipole-W-shaped, and W-shaped-dipole solitons, we see that they are completely determined by the fiber
parameters (e.g., group-velocity dispersion coefficient, cubic nonlinearity
coefficient). This indicates that these localized waveforms can be formed in
birefringent fibers of different characteristics.

\section{Periodic wave trains}

We present here a broad class of exact analytical periodic wave solutions
for the coupled NLS equations (\ref{4}) and (\ref{5}) in the strong
birefringent case $\sigma =1$. These nonlinear waves are of practical
relevance as they serve as a model of pulse train propagation in optics
fibers\ \cite{D}.

\begin{description}
\item[\textbf{TYPE-I}] 
\end{description}

We find that the amplitude equations (\ref{13}) and (\ref{14}) satisfy the
periodic solutions,%
\begin{equation}
u\left( x\right) =\lambda \,\mathrm{cn}\left( w(x-\eta ),k\right) \mathrm{sn}%
\left( w(x-\eta ),k\right) ,  \label{64}
\end{equation}%
\begin{equation}
v\left( x\right) =\rho \,\mathrm{dn}\left( w(x-\eta ),k\right) \mathrm{cn}%
\left( w(x-\eta ),k\right) ,  \label{65}
\end{equation}%
where%
\begin{equation}
w^{2}=-\frac{c}{5k^{2}-4},\quad \lambda ^{2}=-\frac{6ck^{4}}{a(5k^{2}-4)}%
,\quad \rho ^{2}=\frac{\lambda ^{2}}{k^{2}},\quad d=\frac{(5k^{2}-1)c}{%
5k^{2}-4},  \label{66}
\end{equation}

\noindent and $\eta $ being the position of the nonlinear wave at $z=0$.
Here in the solutions (\ref{64}) and (\ref{65}), $\mathrm{cn,}$ $\mathrm{sn}$
and $\mathrm{dn}$ are Jacobi elliptic functions of modulus $k$\ which is in
the range $0<k<1$.

Then, using Eqs. (\ref{15}) and (\ref{66}), one obtains the wave parameters $%
w$, $\lambda $ and $\kappa $ as 
\begin{equation}
w=\sqrt{-\frac{4\alpha }{3\beta _{2}}},\quad \lambda =\pm k^{2}\sqrt{\frac{%
4\alpha }{\gamma }},\quad \kappa =\frac{5\alpha (2k^{2}-1)}{3}+\frac{\beta
_{2}\delta ^{2}}{2},  \label{67}
\end{equation}%
where $\delta$ is free parameter. Note that we have for positive and
negative value of parameter $\lambda $ two different solution for the
amplitude $\rho $ as 
\begin{equation}
\rho =k\sqrt{\frac{4\alpha }{\gamma }},\quad \rho =-k\sqrt{\frac{4\alpha }{%
\gamma }},  \label{68}
\end{equation}%
because of relation $\rho ^{2}=\lambda ^{2}/k^{2}$. Insertion of these
results into Eqs. (\ref{6}) and (\ref{7}), we get a class of exact periodic
wave solutions for the coupled NLS system (\ref{4}) and (\ref{5}) as
\begin{equation}
U(z,\tau )=\lambda \,\mathrm{cn}\left( w(\tau -qz-\eta ),k\right) \mathrm{sn}%
\left( w(\tau -qz-\eta ),k\right) \exp [i(\kappa z-\delta \tau +\theta )],
\label{69}
\end{equation}

\begin{equation}
V(z,\tau )=\rho \,\mathrm{dn}\left( w(\tau -qz-\eta ),k\right) \mathrm{cn}%
\left( w(\tau -qz-\eta ),k\right) \exp [i(\kappa z-\delta \tau +\theta )],
\label{70}
\end{equation}

\noindent which exist for waveguide parameters obeying the physical
conditions $\alpha \beta _{2}<0$ and $\alpha \gamma >0$.

\begin{description}
\item[\textbf{TYPE-II}] 
\end{description}

We find a family of exact periodic wave solutions for the amplitude
equations (\ref{13}) and (\ref{14})\ as 
\begin{equation}
u\left( x\right) =R\,\mathrm{dn}\left( \mu (x-\eta ),k\right) \mathrm{cn}%
\left( \mu (x-\eta ),k\right) ,  \label{71}
\end{equation}%
\begin{equation}
v\left( x\right) =P\,\mathrm{cn}\left( \mu (x-\eta ),k\right) \mathrm{sn}%
\left( \mu (x-\eta ),k\right) ,  \label{72}
\end{equation}%
where%
\begin{equation}
\mu ^{2}=-\frac{d}{5k^{2}-4},\quad P^{2}=-\frac{6dk^{4}}{a(5k^{2}-4)},\quad
R^{2}=\frac{P^{2}}{k^{2}},\quad d=\frac{(5k^{2}-4)c}{5k^{2}-1}.  \label{73}
\end{equation}

\noindent Equations (\ref{15}) and (\ref{73}) yield the soliton parameters:%
\begin{equation}
\mu \,=\sqrt{\frac{4\alpha }{3\beta _{2}}},\quad P\,=\pm k^{2}\sqrt{-\frac{%
4\alpha }{\gamma }},\quad \kappa =-\frac{5\alpha (2k^{2}-1)}{3}+\frac{\beta
_{2}\delta ^{2}}{2},  \label{74}
\end{equation}%
where $\delta$ is free parameter. We have for positive and negative value of
parameter $P$ two different solution for the amplitude $R$ as 
\begin{equation}
R\,=k\sqrt{-\frac{4\alpha }{\gamma }},\quad R\,=-k\sqrt{-\frac{4\alpha }{%
\gamma }},  \label{75}
\end{equation}%
because of relation $R^{2}=P\,^{2}/k^{2}$. \noindent If we insert these
results into Eqs. (\ref{6}) and (\ref{7}), one obtains exact analytical
soliton solution for the coupled NLS equations (\ref{4}) and (\ref{5}) as
\begin{equation}
U(z,\tau )=R\,\mathrm{dn}\left( \mu (\tau -qz-\eta ),k\right) \mathrm{cn}%
\left( \mu (\tau -qz-\eta ),k\right) \exp [i(\kappa z-\delta \tau +\theta )],
\label{76}
\end{equation}

\begin{equation}
V(z,\tau )=P\,\mathrm{cn}\left( \mu (\tau -qz-\eta ),k\right) \mathrm{sn}%
\left( \mu (\tau -qz-\eta ),k\right) \exp [i(\kappa z-\delta \tau +\theta )].
\label{77}
\end{equation}

\noindent One sees that the existence conditions of these periodic waves are 
$\alpha \beta _{2}>0$ and $\alpha \gamma <0,$ which differ from those given
in Eqs. (\ref{69}) and (\ref{70}).

\begin{description}
\item[\textbf{TYPE-III}] 
\end{description}

We find that Eqs. (\ref{13}) and (\ref{14}) admit exact analytical periodic
wave solutions of the form,%
\begin{equation}
u\left( x\right) =S\,\mathrm{cn}\left( \sigma (x-\eta ),k\right) ,
\label{78}
\end{equation}%
\begin{equation}
v\left( x\right) =Q\,\mathrm{sn}\left( \sigma (x-\eta ),k\right) ,
\label{79}
\end{equation}%
where%
\begin{equation}
\sigma ^{2}=\frac{d-c}{k^{2}},\quad S^{2}=\frac{d-c(k^{2}+1)}{ak^{2}},\quad
Q^{2}=\frac{d(1-2k^{2})+c(k^{2}-1)}{ak^{2}}.  \label{80}
\end{equation}

\noindent With use of Eqs. (\ref{15}) and (\ref{80}), the parameters $\sigma $ and $S$ become 
\begin{equation}
\sigma =\sqrt{\frac{4\alpha }{\beta _{2}k^{2}}},  \label{81}
\end{equation}%
\begin{equation}
S=\pm \sqrt{\frac{2\kappa k^{2}-\beta _{2}\delta ^{2}k^{2}-2\alpha (k^{2}+2)%
}{2\gamma k^{2}}}.  \label{82}
\end{equation}%
We have for positive and negative value of parameter $S$ two different
solution for the amplitude $Q$ as 
\begin{equation}
Q=\sqrt{\frac{2\kappa k^{2}-\beta _{2}\delta ^{2}k^{2}+2\alpha (3k^{2}-2)}{%
2\gamma k^{2}}},\quad Q=-\sqrt{\frac{2\kappa k^{2}-\beta _{2}\delta
^{2}k^{2}+2\alpha (3k^{2}-2)}{2\gamma k^{2}}}.  \label{83}
\end{equation}%
Further substitution of these closed form solutions into Eqs. (\ref{6}) and (\ref{7}) leads to the following exact periodic solution for the coupled NLS
system (\ref{4}) and (\ref{5}):%
\begin{equation}
U(z,\tau )=S\,\mathrm{cn}\left( \sigma (\tau -qz-\eta ),k\right) \exp
[i(\kappa z-\delta \tau +\theta )],  \label{84}
\end{equation}%
\begin{equation}
V(z,\tau )=Q\,\mathrm{sn}\left( \sigma (\tau -qz-\eta ),k\right) \exp
[i(\kappa z-\delta \tau +\theta )].  \label{85}
\end{equation}
Note that these nonlinear periodic solutions are dependent on two free
parameters $\kappa $ and $\delta $ and exist for $\alpha \beta _{2}>0$, $%
\gamma \left[ 2\kappa k^{2}-\beta _{2}\delta ^{2}k^{2}-2\alpha (k^{2}+2)%
\right] >0$ and $\gamma \left[ 2\kappa k^{2}-\beta _{2}\delta
^{2}k^{2}+2\alpha (3k^{2}-2)\right] >0$.

\begin{description}
\item[\textbf{TYPE-IV}] 
\end{description}

We can get a class of exact analytical soliton solutions for Eqs. (\ref{13})
and (\ref{14}) as
\begin{equation}
u\left( x\right) =f\,\,\mathrm{sn}\left( r(x-\eta ),k\right) ,  \label{86}
\end{equation}%
\begin{equation}
v\left( x\right) =g\,\,\mathrm{cn}\left( r(x-\eta ),k\right) ,  \label{87}
\end{equation}%
where%
\begin{equation}
r^{2}=\frac{c-d}{k^{2}},\quad f^{2}=\frac{c(1-2k^{2})+d(k^{2}-1)}{ak^{2}}%
,\quad g^{2}=\frac{c-d(k^{2}+1)}{ak^{2}}.  \label{88}
\end{equation}

\noindent Thus, using Eqs. (\ref{15}) and (\ref{88}), we obtain the waver
parameters $r$ and $f$ as 
\begin{equation}
r=\sqrt{-\frac{4\alpha }{\beta _{2}k^{2}}},  \label{89}
\end{equation}%
\begin{equation}
f=\pm \sqrt{\frac{2\kappa k^{2}-\beta _{2}\delta ^{2}k^{2}-2\alpha (3k^{2}-2)%
}{2\gamma k^{2}}}.  \label{90}
\end{equation}%
We have for positive and negative value of parameter $f$ two different
solution for the amplitude $g$ as 
\begin{equation}
g=\sqrt{\frac{2\kappa k^{2}-\beta _{2}\delta ^{2}k^{2}+2\alpha (k^{2}+2)}{%
2\gamma k^{2}}},\quad g=-\sqrt{\frac{2\kappa k^{2}-\beta _{2}\delta
^{2}k^{2}+2\alpha (k^{2}+2)}{2\gamma k^{2}}}.  \label{91}
\end{equation}%
Hence, the exact periodic wave solutions of the coupled NLS equations (\ref{4}) and (\ref{5}) can be written as
\begin{equation}
U(z,\tau )=f\,\,\,\mathrm{sn}\left( r(\tau -qz-\eta ),k\right) \exp
[i(\kappa z-\delta \tau +\theta )],  \label{92}
\end{equation}%
\begin{equation}
V(z,\tau )=g\,\,\mathrm{cn}\left( r(\tau -qz-\eta ),k\right) \exp [i(\kappa
z-\delta \tau +\theta )].  \label{93}
\end{equation}%
Note that these nonlinear periodic solutions are dependent on two free
parameters $\kappa $ and $\delta $ and these periodic waves do exist
provided that $\alpha \beta _{2}<0, $ $\gamma \left[ 2\kappa k^{2}-\beta
_{2}\delta ^{2}k^{2}-2\alpha (3k^{2}-2)\right] >0$ and $\gamma \left[
2\kappa k^{2}-\beta _{2}\delta ^{2}k^{2}+2\alpha (k^{2}+2)\right] >0$.

\begin{description}
\item[\textbf{TYPE-V}] 
\end{description}

We obtained another class of exact analytical periodic solutions for Eqs. (\ref{13}) and (\ref{14}) as 
\begin{equation}
u(x)=A-B\,\mathrm{cn}^{2}\left( \epsilon (x-\eta ),k\right) ,  \label{94}
\end{equation}%
\begin{equation}
v(x)=F\,\mathrm{cn}\left( \epsilon (x-\eta ),k\right) \mathrm{sn}\left(
\epsilon (x-\eta ),k\right) .  \label{95}
\end{equation}%
The parameters in these solutions are given by Eqs. (\ref{A1})-(\ref{A5})
presented in Appendix A. We found in this Appendix A the relation $d=f(k)c$
where the function $f(k)$ can be found by cubic Eq. (\ref{A22}). However,
this cubic equation has three different solutions for the function $f(k)$.
We choose the solution for function $f(k)$ which leads to the value $%
f(1)=7/8 $ for $k=1$. This condition follows from Eq. (\ref{49}) because the
Type-V periodic solutions reduce to W-shaped-dipole solitons in the limiting
case with $k=1$. Using the relation $d=f(k)c$ found in Appendix A and the
wave number given in (\ref{A29}) as 
\begin{equation}
\kappa =-\frac{\alpha (1+f)}{1-f}+\frac{1}{2}\beta _{2}\delta ^{2},
\label{96}
\end{equation}%
where $\delta $ is free parameter. We have by (\ref{15}) the parameters $c$
and $d$: 
\begin{equation}
c=-\frac{4\alpha }{\beta _{2}(1-f)},~~~~d=-\frac{4\alpha f}{\beta _{2}(1-f)}.
\label{97}
\end{equation}%
The parameter $\epsilon $ follows from Eq. (\ref{A18}) as 
\begin{equation}
\epsilon =\left( \frac{f^{2}c(24k^{2}-30)+fc(-35k^{2}+40)+c(11k^{2}-10)}{%
f(-8k^{4}+110k^{2}-80)+(17k^{4}-110k^{2}+80)}\right) ^{1/2}.  \label{98}
\end{equation}%
Note that without loss of generality, we defined by Eq. (\ref{98}) the
positive value for parameter $\epsilon $. The amplitudes for solutions in
Eqs. (\ref{94}) and (\ref{95}) follow by (\ref{A5}) and (\ref{A6}) as 
\begin{equation}
A=\pm \sqrt{\frac{\epsilon ^{2}(4-5k^{2})-fc}{a}},\quad B=\pm \sqrt{\frac{%
8\epsilon ^{2}(2-k^{2})+2c(1-3f)}{a}},  \label{99}
\end{equation}%
\begin{equation}
F=\pm \sqrt{\frac{8\epsilon ^{2}(2-k^{2})+2c(1-3f)}{a}}.  \label{100}
\end{equation}%
We note that the signs in Eqs. (\ref{99}) and (\ref{100}) can be chosen
independent on signs of other amplitudes. Hence, one can choose eight
different sequences of signs in (\ref{99}) and (\ref{100}). This means we
have eight different solutions given by Eqs. (\ref{94}) and (\ref{95}). The
insertion of these results into Eqs. (\ref{6}) and (\ref{7}) yields eight
soliton pulse solutions for the coupled NLS equations (\ref{4}) and (\ref{5}) as 
\begin{equation}
U(z,t)=\left\{ A-B\,\mathrm{cn}^{2}\left( \epsilon (\tau -qz-\eta ),k\right)
\right\} \exp [i(\kappa z-\delta \tau +\theta )],  \label{101}
\end{equation}%
\begin{equation}
V(z,t)=F\,\mathrm{cn}\left( \epsilon (\tau -qz-\eta ),k\right) \mathrm{sn}%
\left( \epsilon (\tau -qz-\eta ),k\right) \exp [i(\kappa z-\delta \tau
+\theta )],  \label{102}
\end{equation}%
which exist provided that $a[\epsilon ^{2}(4-5k^{2})-fc]>0$, $a[8\epsilon
^{2}(2-k^{2})+2c(1-3f)]>0$ and parameter $\epsilon $ defined in (\ref{98})
is a real number.

\begin{description}
\item[\textbf{TYPE-VI}] 
\end{description}

We find another class of exact analytical periodic solutions for Eqs. (\ref{13}) and (\ref{14}) as 
\begin{equation}
u\left( x\right) =\Lambda\,\mathrm{cn}\left( \nu (x-\eta ),k\right) \mathrm{%
sn}\left( \nu (x-\eta ),k\right) ,  \label{103}
\end{equation}%
\begin{equation}
v\left( x\right) =D-\Gamma\,\mathrm{cn}^{2}\left( \nu (x-\eta),k\right) .
\label{104}
\end{equation}%
The parameters in these solutions are given by Eqs. (\ref{A1})-(\ref{A5})
presented in Appendix A with the change for parameters $A$, $B$, $F$, $%
\epsilon$ to $D$, $\Gamma$, $\Lambda$ and $\nu$ respectively. Moreover, we
should make the change $u$, $v$, $c$ and $d$ to $v$, $u$, $d$ and $c$ in
appropriate equations. This simple principle follows from the symmetry of
Eqs. (\ref{13}) and (\ref{14}). In this case we have the relation $c=f(k)d$
where the function $f(k)$ can be found by cubic Eq. (\ref{A22}). However,
this cubic equation has three different solutions for the function $f(k)$.
We choose the solution for function $f(k)$ which leads to the value $%
f(1)=7/8 $ for $k=1$. This condition follows from Eq. (\ref{58}) because the
Type-VI periodic solutions reduce to dipole-W-shaped solitons in the
limiting case with $k=1$. Using the relation $c=f(k)d$ we have with Eq. (\ref{15}) the wave number as 
\begin{equation}
\kappa=\frac{\alpha(1+f)}{1-f}+\frac{1}{2}\beta_{2}\delta^{2},  \label{105}
\end{equation}%
where $\delta $ is free parameter. In this case the parameters $c$ and $d$ are 
\begin{equation}
c=\frac{4\alpha f}{\beta_{2}(1-f)},~~~~d=\frac{4\alpha}{\beta_{2}(1-f)}.
\label{106}
\end{equation}%
The parameter $\nu$ has the following form, 
\begin{equation}
\nu=\left(\frac{f^{2}d(24k^{2}-30)+fd(-35k^{2}+40)+d(11k^{2}-10)}{%
f(-8k^{4}+110k^{2}-80)+(17k^{4}-110k^{2}+80)}\right)^{1/2}.  \label{107}
\end{equation}
We can define here the positive value for parameter $\nu$. The amplitudes
for solutions in Eqs. (\ref{103}) and (\ref{104}) have the following form, 
\begin{equation}
D=\pm \sqrt{\frac{\nu^{2}(4-5k^{2})-fd}{a}},\quad \Gamma=\pm\sqrt{\frac{%
8\nu^{2}(2-k^{2})+2d(1-3f)}{a}},  \label{108}
\end{equation}%
\begin{equation}
\Lambda=\pm\sqrt{\frac{8\nu^{2}(2-k^{2})+2d(1-3f)}{a}}.  \label{109}
\end{equation}%
We note that the signs in Eqs. (\ref{108}) and (\ref{109}) can be chosen
independent on signs of other amplitudes. Hence, one can choose eight
different sequences of signs in (\ref{108}) and (\ref{109}). This means we
have eight different solutions given by Eqs. (\ref{103}) and (\ref{104}).
The insertion of these results into Eqs. (\ref{6}) and (\ref{7}) yields
eight soliton pulse solutions for the coupled NLS equations (\ref{4}) and (\ref{5}) as 
\begin{equation}
U(z,\tau )=\Lambda \,\mathrm{cn}\left( \nu (\tau -qz-\eta ),k\right)\mathrm{%
sn}\left( \nu (\tau -qz-\eta ),k\right) \exp [i(\kappa z-\delta\tau +\theta
)],  \label{110}
\end{equation}%
\begin{equation}
V(z,\tau )=\left\{ D-\Gamma \,\mathrm{cn}^{2}\left( \nu (\tau-qz-\eta
),k\right) \right\} \exp [i(\kappa z-\delta \tau +\theta )],  \label{111}
\end{equation}

\noindent which exist provided that $a[\nu^{2}(4-5k^{2})-fd]>0$, $a[8\nu
^{2}(2-k^{2})+2d(1-3f)]>0$ and parameter $\nu $ defined in (\ref{107}) is a real number.

\begin{figure}[h]
\includegraphics[width=1\textwidth]{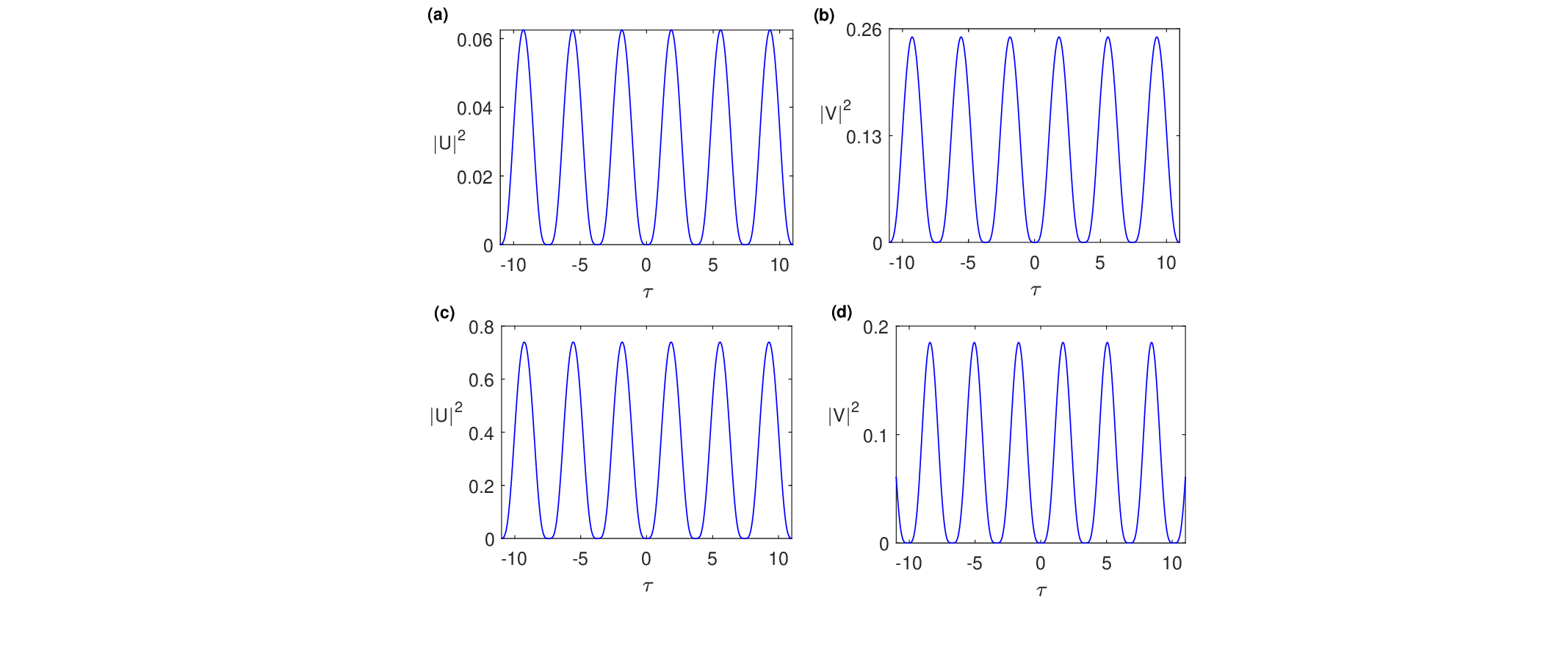}
\caption{The intensity versus $\protect\tau $ for the (a) periodic solution (%
\protect\ref{69}) (b) periodic solution (\protect\ref{70}) (c) periodic
solution (\protect\ref{76}) and (d) periodic solution (\protect\ref{69}) for
the values mentioned in the text.}
\label{FIG.4.}
\end{figure}

As the intensity profile of periodic solutions exhibits an oscillating
character, we only take the types I and II as examples to present the
intensity of above periodic waves. Figures 4(a)-(b) depict the intensity
wave profile of periodic waves (\ref{69}) and (\ref{70}) for the parameters
values: $\beta _{2}=-1,\alpha =0.75,$ $\gamma =1,$ and $\delta =-0.01.$ The
results for the periodic waves (\ref{76}) and (\ref{77}) are illustrated in
Figs. 4(c)-(d) for the values $\beta _{2}=-1,\alpha =-0.75,\gamma =1,$ and $%
\delta =0.01.$ Here the position $\eta $ of periodic wave solutions at $z=0$ is chosen to be equal to zero and the value of the elliptic modulus $k$
is taken as $k=0.5$. As seen these nonlinear waves present the periodic
property, which makes them a setting of light pulse train propagation in
nonlinear fiber systems.

Further analysis of the obtained periodic wave solutions indicates that for
special case of modulus $k=1$ that corresponds to the long wave limit, the
various soliton solutions presented above can be readily obtained. We note
that depending on the choice of the modulus $k$ that may take any value from 
$0<k<1$, we can obtain periodic wave pairs with different values of
amplitudes and widths. These results could offer the possibility to realize
rich soliton structures and periodic wave trains experimentally in
birefringent nonlinear fiber systems.

\section{Conclusion}

In conclusion, we have analyzed the existence of localized and periodic
waves in an homogeneous birefringent optical fiber with Kerr nonlinear
response. We have found that the fiber system allows the propagation of a
rich variety of envelope solitons including dipole-bright, bright-dipole,
bright-dark, dark-bright, W-shaped-dipole and dipole-W-shaped types. These
results indicate that in addition to single hump solitons, the transmitting
birefringent medium could also display multihump solitons having the form of
dipole and W-shaped type solitons. We have found that these solitons can be
established in both the normal and anomalous group velocity dispersion
regions. We also showed that nonlinear periodic waves of different forms
could also exist in the transmiting medium. Thus these results confirm the
existence of both single and multiple-humped solitons as well as a diversity
of periodic waves in a birefringent fiber media, which could open up novel
applications in optical communications.

\appendix:

\section{Solution of Type-V}

In this appendix, we provide the detailed calculations for obtaining all
parameters for solution of Type-V. We also found here the function $f(k)$ in
relation $d=f(k)c$ and explicit form for the wave number $\kappa$. We have
the following algebraic equations for solutions of Type-V: 
\begin{equation}
2B\epsilon ^{2}(k^{2}-1)+aA^{3}+cA=0,  \label{A1}
\end{equation}%
\begin{equation}
4\epsilon ^{2}(1-2k^{2})-3aA^{2}+aAB-c=0,  \label{A2}
\end{equation}%
\begin{equation}
6\epsilon ^{2}k^{2}+2aAB-aB^{2}=0,  \label{A3}
\end{equation}%
\begin{equation}
\epsilon ^{2}(5k^{2}-4)+aA^{2}+d=0,  \label{A4}
\end{equation}%
\begin{equation}
B^{2}=F^{2}.  \label{A5}
\end{equation}%
Using Eqs. (\ref{A2})-(\ref{A4}), we obtain 
\begin{equation}
aB^{2}=8\epsilon ^{2}(2-k^{2})+2c-6d,~~~~aA^{2}=\epsilon ^{2}(4-5k^{2})-d.
\label{A6}
\end{equation}%
Moreover, using Eqs. (\ref{A1}), (\ref{A4}), we get 
\begin{equation}
2aAB\epsilon ^{2}(k^{2}-1)+\epsilon ^{4}(5k^{2}-4)^{2}+\epsilon
^{2}[d(10k^{2}-8)-c(5k^{2}-4)]+d^{2}-cd=0.  \label{A7}
\end{equation}%
In addition, from Eqs. (\ref{A2}) and (\ref{A4}), we have 
\begin{equation}
aAB=\epsilon ^{2}(8-7k^{2})-3d+c.  \label{A8}
\end{equation}%
Also, using Eqs. (\ref{A7}) and (\ref{A8}), we find 
\begin{equation}
\epsilon ^{4}(11k^{4}-10k^{2})+[d(4k^{2}-2)+c(2-3k^{2})]\epsilon
^{2}+d^{2}-cd=0.  \label{A9}
\end{equation}%
Now using Eq. (\ref{A1}), we have 
\begin{equation}
2aB^{2}\epsilon ^{2}(k^{2}-1)+aA^{2}aAB+caAB=0.  \label{A10}
\end{equation}%
Using Eqs. (\ref{A10}), (\ref{A6}) and (\ref{A8}), one gets 
\begin{equation}
\epsilon ^{4}(19k^{4}-20k^{2})+[d(10k^{2}-8)+c(8-8k^{2})]\epsilon
^{2}+3d^{2}+c^{2}-4cd=0.  \label{A11}
\end{equation}%
Then, the system of Eqs. (\ref{A9}) and (\ref{A11}) can be written as 
\begin{equation}
\epsilon ^{4}+\left( \frac{d(4k^{2}-2)+c(2-3k^{2})}{11k^{4}-10k^{2}}\right)
\epsilon ^{2}+\frac{d^{2}-cd}{11k^{4}-10k^{2}}=0,  \label{A12}
\end{equation}%
\begin{equation}
\epsilon ^{4}+\left( \frac{d(10k^{2}-8)+c(8-8k^{2})}{19k^{4}-20k^{2}}\right)
\epsilon ^{2}+\frac{3d^{2}+c^{2}-4cd}{19k^{4}-20k^{2}}=0.  \label{A13}
\end{equation}%
Now using Eq. (\ref{A8}) we have 
\begin{equation}
a^{2}A^{2}B^{2}=[\epsilon ^{2}(8-7k^{2})+(c-3d)]^{2}.  \label{A14}
\end{equation}%
Using Eq. (\ref{A6}) one gets 
\begin{equation}
a^{2}A^{2}B^{2}=[\epsilon ^{2}(4-5k^{2})-d][\epsilon
^{2}(16-8k^{2})+(2c-6d)].  \label{A15}
\end{equation}%
Using Eqs. (\ref{A14}) and (\ref{A15}), we find 
\begin{equation}
\epsilon ^{4}+\left( \frac{d(4k^{2}-8)+c(8-4k^{2})}{9k^{4}}\right) \epsilon
^{2}+\frac{3d^{2}+c^{2}-4cd}{9k^{4}}=0.  \label{A16}
\end{equation}%
Subtraction of Eq. (\ref{A16}) from Eq. (\ref{A12}), we have 
\begin{equation}
\left( \frac{d(4k^{2}-2)+c(2-3k^{2})}{11k^{2}-10}-\frac{%
d(4k^{2}-8)+c(8-4k^{2})}{9k^{2}}\right) \epsilon ^{2}+\frac{d^{2}-cd}{%
11k^{2}-10}-\frac{3d^{2}+c^{2}-4cd}{9k^{2}}=0.  \label{A17}
\end{equation}%
Thus, Eq. (\ref{A17}) yields 
\begin{equation}
\epsilon ^{2}=\frac{d^{2}(24k^{2}-30)+cd(-35k^{2}+40)+c^{2}(11k^{2}-10)}{%
d(-8k^{4}+110k^{2}-80)+c(17k^{4}-110k^{2}+80)}.  \label{A18}
\end{equation}
Subtraction of Eq. (\ref{A16}) from Eq. (\ref{A13}) yields 
\begin{equation}
\left( \frac{d(10k^{2}-8)+c(8-8k^{2})}{19k^{2}-20}-\frac{%
d(4k^{2}-8)+c(8-4k^{2})}{9k^{2}}\right) \epsilon ^{2}+\frac{3d^{2}+c^{2}-4cd%
}{19k^{2}-20}-\frac{3d^{2}+c^{2}-4cd}{9k^{2}}=0.  \label{A19}
\end{equation}%
Moreover, Eq. (\ref{A19}) yields 
\begin{equation}
\epsilon ^{2}=\frac{d^{2}(30k^{2}-60)+cd(-40k^{2}+80)+c^{2}(10k^{2}-20)}{%
d(14k^{4}+160k^{2}-160)+c(4k^{4}-160k^{2}+160)}.  \label{A20}
\end{equation}%
Hence Eqs. (\ref{A18}) and (\ref{A20}) with $d=f(k)c$ leads to the cubic
equation for $f(k)$: 
\begin{equation}
\frac{f^{2}(24k^{2}-30)+f(-35k^{2}+40)+(11k^{2}-10)}{%
f(-8k^{4}+110k^{2}-80)+(17k^{4}-110k^{2}+80)}=\frac{%
f^{2}(30k^{2}-60)+f(-40k^{2}+80)+(10k^{2}-20)}{%
f(14k^{4}+160k^{2}-160)+(4k^{4}-160k^{2}+160)}.  \label{A21}
\end{equation}%
Equation (\ref{A21}) can be written in standard cubic form as 
\begin{equation}
G_{3}(k)f^{3}+G_{2}(k)f^{2}+G_{1}(k)f+G_{0}(k)=0,  \label{A22}
\end{equation}%
where the polynomial $G_{j}(k)$ are 
\begin{equation}
G_{3}(k)=(24k^{2}-30)P_{3}(k)-(30k^{2}-60)P_{1}(k),  \label{A23}
\end{equation}%
\begin{equation}
G_{2}(k)=(-35k^{2}+40)P_{3}(k)+(24k^{2}-30)P_{4}(k)-(-40k^{2}+80)P_{1}(k)-(30k^{2}-60)P_{2}(k),
\label{A24}
\end{equation}%
\begin{equation}
G_{1}(k)=(11k^{2}-10)P_{3}(k)+(-35k^{2}+40)P_{4}(k)-(10k^{2}-20)P_{1}(k)-(-40k^{2}+80)P_{2}(k),
\label{A25}
\end{equation}%
\begin{equation}
G_{0}(k)=(11k^{2}-10)P_{4}(k)-(10k^{2}-20)P_{2}(k).  \label{A26}
\end{equation}%
Here the polynomials $P_{j}(k)$ are given as 
\begin{equation}
P_{1}(k)=-8k^{4}+110k^{2}-80,~~~~P_{2}(k)=17k^{4}-110k^{2}+80,  \label{A27}
\end{equation}%
\begin{equation}
P_{3}(k)=14k^{4}+160k^{2}-160,~~~~P_{4}(k)=4k^{4}-160k^{2}+160.  \label{A28}
\end{equation}%
The relation $d=f(k)c$ with Eq. (\ref{15}) yield the wave number $\kappa$ as 
\begin{equation}
\kappa=-\frac{\alpha(1+f)}{1-f}+\frac{1}{2}\beta_{2}\delta^{2}.  \label{A29}
\end{equation}%
Let for example $k=1$ then Eq. (\ref{A21}) or cubic Eq. (\ref{A22}) can be
written as 
\begin{equation}
(-6f^{2}+5f+1)(14f+4)-(-30f^{2}+40f-10)(22f-13)=0.  \label{A30}
\end{equation}%
This equation has the solution $f=7/8,$ leading to the relationship $d=7c/8$
obtained for the case of W-shaped-dipole solitons. We note that in general
case the cubic Eq. (\ref{A22}) has three different solutions. We choose the
solution for function $f(k)$ which leads to the value $f(1)=7/8$ for $k=1$.
This condition follows from Eq. (\ref{49}) because the Type-VI periodic
solutions reduce to dipole-W-shaped solitons in the limiting case with $k=1$.

\end{document}